\documentclass[prd,amsmath,amssymb,superscriptaddress,nofootinbib,twocolumn]{revtex4}
\usepackage{amsfonts}
\usepackage{multirow}
\usepackage{makecell}
\usepackage{mathrsfs}
\usepackage{graphicx}
\usepackage{amsmath}
\usepackage{amssymb}
\usepackage{bm}
\usepackage{bbm}
\usepackage{color}
\usepackage{ulem}
\usepackage{array}
\usepackage{diagbox}
\usepackage{cancel}
\usepackage{float}
\usepackage{subcaption}

\newcommand{\nn}{\nonumber}

\newcommand{\beq}{\begin{equation}}
\newcommand{\eeq}{\end{equation}}
\newcommand{\bqa}{\begin{eqnarray}}
\newcommand{\eqa}{\end{eqnarray}}

\newcommand{\bseq}{\begin{subequations}}
\newcommand{\eseq}{\end{subequations}}

\makeatletter

\begin{document}

\title{Deeply virtual pion production through two-loop order}

\author{Wen Chen~\footnote{chenwenphy@gmail.com}}
\affiliation{State Key Laboratory of Nuclear Physics and Technology, Institute of Quantum Matter, South China Normal University, Guangzhou 510006, China\vspace{0.2cm}}
\affiliation{Guangdong Basic Research Center of Excellence for Structure and Fundamental Interactions of Matter, Guangdong Provincial Key Laboratory of Nuclear Science, Guangzhou 510006, China\vspace{0.2cm}}
\author{Feng Feng~\footnote{f.feng@outlook.com}}
\affiliation{China University of Mining and Technology, Beijing 100083, China\vspace{0.2cm}}
\affiliation{Institute of High Energy Physics, Chinese Academy of Sciences, Beijing 100049, China\vspace{0.2cm}}
\author{Yu Jia~\footnote{yjia@m.scnu.edu.cn}  }
\affiliation{State Key Laboratory of Nuclear Physics and Technology, Institute of Quantum Matter, South China Normal University, Guangzhou 510006, China\vspace{0.2cm}}
\affiliation{Guangdong Basic Research Center of Excellence for Structure and Fundamental Interactions of Matter, Guangdong Provincial Key Laboratory of Nuclear Science, Guangzhou 510006, China\vspace{0.2cm}}
\author{Qin-Tao Song~\footnote{songqintao@zzu.edu.cn}}
\affiliation{School of Physics, Zhengzhou University, Zhengzhou, Henan 450001, China\vspace{0.2cm}}
\author{Guang Tang~\footnote{tangg@ihep.ac.cn}}
\affiliation{Institute of High Energy Physics, Chinese Academy of Sciences, Beijing 100049, China\vspace{0.2cm}}
\affiliation{School of Physical Sciences,
University of Chinese Academy of Sciences, Beijing 100049, China\vspace{0.2cm}}
\author{Zhe-Yu Wang~\footnote{wangzheyu@ihep.ac.cn}}
\affiliation{Institute of High Energy Physics, Chinese Academy of Sciences, Beijing 100049, China\vspace{0.2cm}}
\affiliation{School of Physical Sciences,
University of Chinese Academy of Sciences, Beijing 100049, China\vspace{0.2cm}}
\date{\today}

\begin{abstract}

Deeply virtual meson production (DVMP) is among the most prominent channels to extract
the nucleon's generalized parton distributions (GPDs) at $ep$ scattering facilities such as {\tt JLab} and the upcoming {\tt EIC/EicC} experiments, which plays a vital role in unravelling the three-dimensional internal structure of nucleon.
In this work we calculate for the first time the next-to-next-to-leading order (NNLO) QCD radiative corrections 
to the DV$\pi$P processes $\gamma^* p\to \pi^+ n$ and $\gamma^* p\to \pi^0 p$ in the generalized Bjorken limit
$Q^2\gg \vert t\vert,  \Lambda_{\text{QCD}}^2$, accurate at the leading twist within collinear factorization framework.
The impact of the two-loop QCD corrections appears to be positive and substantial, 
including which considerably improves the agreement between the perturbative QCD prediction and the available {\tt JLab} data. 
In addition to the differential longitudinal DV$\pi$P cross section, we also study the impact of the two-loop QCD corrections on the transverse single-spin asymmetries (TSSA) in some benchmark kinematics at {\tt JLab}, {\tt EIC} and {\tt EicC}. 

\end{abstract}

\maketitle


\noindent{\color{blue}\it Introduction.} Study of generalized parton distributions (GPDs) of nucleons~\cite{Muller:1994ses,Ji:1996nm,Radyushkin:1996nd} stands as one of the central goals of contemporary hadron physics.
By correlating partonic longitudinal momentum with transverse spatial position~\cite{Burkardt:2000za}, GPDs provide a 
unique tool for accessing three-dimensional partonic structure of nucleons. In addition, GPDs are intimately linked with 
the proton spin decomposition~\cite{Ji:1996ek,Leader:2013jra,Aidala:2012mv,Ji:2020ena},  gravitational form factors~\cite{Burkert:2023wzr,Kumano:2017lhr,Freese:2019bhb,Sun:2020wfo,Duran:2022xag,Hackett:2023rif,Cao:2024zlf,Guo:2025jiz,Tanaka:2025pny,Hatta:2025ryj}, and internal mechanical distributions~\cite{Polyakov:2002yz,Lorce:2018egm,Polyakov:2018zvc,Burkert:2018bqq,Kumericki:2019ddg,Shanahan:2018nnv,Lorce:2025ayr,Freese:2021czn} (for comprehensive review on various aspects of GPDs, see Refs.~\cite{Diehl:2003ny, Belitsky:2005qn, Goeke:2001tz, Boffi:2007yc}.).

In sharp contrast to the familiar one-dimensional PDFs of proton, to date our theoretical knowledge toward the nucleon GPDs, which are functions of three variables, are still rather limited. 
Some influential parameterized models of nucleon GPDs are available in the market~\cite{Goloskokov:2009ia,Goloskokov:2011rd, Kroll:2012sm,Goloskokov:2007nt,Guo:2025muf,Vanderhaeghen:1999xj,Kumericki:2015lhb,Kumericki:2016ehc,Moutarde:2019tqa,Kriesten:2021sqc}. Recently there also appear the numerical calculation of GPDs of pion and nucleon at some specific value of skewness from lattice QCD simulation~\cite{Bhattacharya:2024wtg,Bhattacharya:2023jsc,Chu:2025kew,Cichy:2023dgk,Alexandrou:2020zbe,Lin:2021brq,Ding:2024saz,Gao:2025inf, Lin:2023gxz, HadStruc:2024rix, Dutrieux:2026grg}. 
On the experimental side,  GPDs can be accessed through the gold-plated hard exclusive processes such as the deeply virtual Compton scattering (DVCS)~\cite{Muller:1994ses,Ji:1996nm,Radyushkin:1996nd} and deeply virtual meson production (DVMP)~\cite{Radyushkin:1996ru,Mankiewicz:1997uy,Mankiewicz:1998kg,Frankfurt:1999xe,Vanderhaeghen:1999xj,Frankfurt:1999fp}. 
Over the past decade, various experimental facilities such as {\tt HERA}, 
{\tt COMPASS} and {\tt JLab} have measured the 
DVCS~\cite{H1:2009wnw,ZEUS:2008hcd,HERMES:2012idp, COMPASS:2018pup,CLAS:2018bgk,CLAS:2018ddh,CLAS:2024qhy} and DVMP~\cite{HERMES:2007hrc,HERMES:2009gtv,CLAS:2008rpm,CLAS:2014jpc,JeffersonLabHallA:2016wye,JeffersonLabHallA:2017hky, JeffersonLabHallA:2020dhq,CLAS:2022iqy,COMPASS:2022xig,
COMPASS:2024hvm, Horn:2007ug}  
in kinematics of $Q^2 \gg |t|, \Lambda_{\text{QCD}}^2$. 
This progress sets the stage for a new exciting era of high-precision tomography at the {\tt JLab} 22 GeV facility \cite{Accardi:2023chb} and  Electron-Ion Colliders in the US ({\tt EIC}) \cite{AbdulKhalek:2021gbh} and China ({\tt EicC}) \cite{Anderle:2021wcy} in future.

Reliable extractions of nucleon GPDs from experimental measurements require a combined analysis of multiple exclusive channels. 
Whereas DVCS probes only flavor-weighted combinations of GPDs, DVMP provides complementary sensitivity that facilitates the separation of individual quark flavor contributions. Consequently, the measurements of DVMP are regarded as an indispensable complements to those of DVCS.

To match the experimental advancements, it is compulsory to provide the most precise predictions for the DVCS and DVMP processes, demanding the inclusion of higher-order $\alpha_s$ corrections.
Very recently the coefficient functions for DVCS~\cite{Braun:2022bpn,Ji:2023xzk,Braun:2025noa} and double deeply virtual Compton scattering~\cite{Braun:2024srt} have been computed at next-to-next-to-leading order (NNLO) within the collinear factorization framework. 
In contrast, phenomenological analyses of DVMP~\cite{Diehl:2007hd,Muller:2013jur,Duplancic:2016bge,Cuic:2023mki} 
is still based on the next-to-leading order (NLO) QCD predictions,  which were first available about two decades ago~\cite{Belitsky:2001nq,Ivanov:2004zv}.
Since the NLO corrections turn out to be sizable for DVMP processes, one naturally wonders whether the NNLO corrections in DVMP may also play an important role in the precise extraction of GPDs.

In this paper, we fill the long-missing gap by presenting the first calculation of the NNLO QCD corrections to a class of 
important DVMP processes, $\pi^+$ and $\pi^0$ production. We find the two-loop corrections are positive and substantial, including which yields a considerably improved description of the available {\tt JLab} data.
This implies that NNLO QCD correction is an indispensable ingredient for a reliable description of the DV$\pi$P cross sections in the upcoming {\tt EIC/EicC} programs and the precise extraction of nucleon GPDs.

\vspace{0.2cm}

\begin{figure}
\centering
\includegraphics[width=0.45\textwidth]{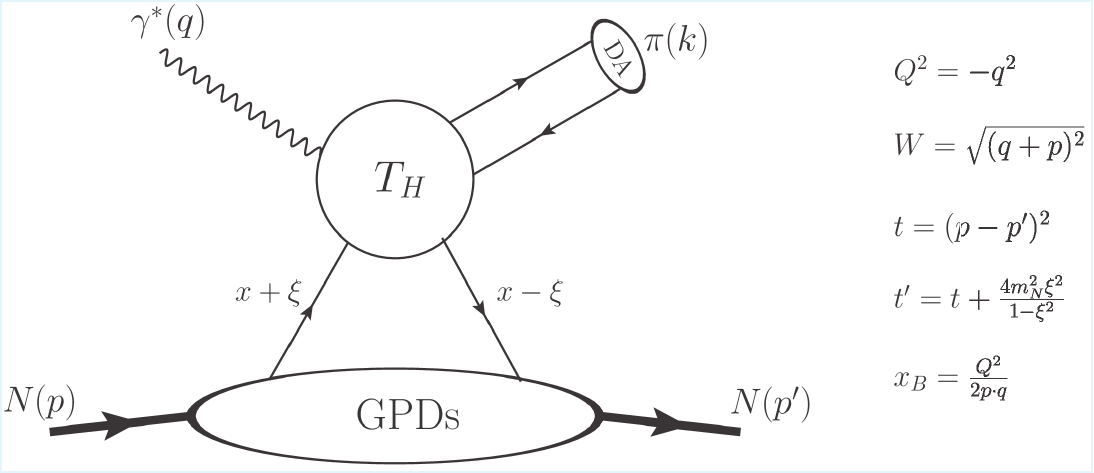}
\caption{A cartoon illustrating the factorized structure of the amplitude for the deeply virtual pion production.}

\label{fig:mp-diagram}
\end{figure}

\noindent{\color{blue}\it Observables related with DV$\pi$P.}
Let us specialize to the DV$\pi$P process 
\beq
\gamma^{\ast}(q,\lambda=0)+ N(p, s)  \to \pi(k) + N(p', s'),
\label{eq:dvmpp}
\eeq
where the momenta and helicities are indicated in parentheses. Concretely speaking, we concentrate on the leading-twist processes, $\gamma_L p\to \pi^+ n$ and $\gamma_L p\to \pi^0 p$,
where the incoming virtual photon is longitudinally polarized~\footnote{The transverse cross section $\sigma_T$
is suppressed by an extra power of $1/Q^2$. At intermediate $Q^2$, 
experimentalists at {\tt JLab} are able to disentangle the $\sigma_L$ from $\sigma_T$ for DV$\pi^+$P
with resort to the Rosenbluth separation technique~\cite{JeffersonLabHallA:2016wye,JeffersonLabHallA:2017hky}.
}. 

The differential longitudinal cross section of DV$\pi$P can be expressed in terms of two helicity amplitudes~\cite{Goloskokov:2009ia,CLAS:2014jpc}:
\beq
\frac{d\sigma_L }{dt} = \frac{  1 }{16 \pi (W^2-m_N^2)\kappa} \left\{  
\left | M_{++}  \right |^{2} + \left | M_{-+}  \right |^{2} \right\},
\label{cro-gk}
\eeq
with $M_{s\,s'}$ signifying helicity amplitudes. The phase space factor is defined as
$\kappa = \sqrt{\Lambda(W^2, -Q^2, m_N^2)}$, with Kallen function 
$\Lambda(a, b, c)=a^2+b^2+c^2-2ab+2ac-2bc$. 

In the collinear factorization regime $Q^2 \gg |t|\, ,\Lambda_{\text{QCD}}^2$, 
the helicity amplitudes in \eqref{cro-gk} can be expressed in terms of the transition FFs (TFFs)~\cite{Goloskokov:2011rd}:
\begin{subequations}
\begin{align}
M_{++}=&\frac{4\pi e f_{\pi}}{N_c Q} \sqrt{1-\xi^2} \left[  \widetilde{ \mathcal{H}}_{\pi} -\frac{\xi^2}{1-\xi^2} \widetilde{ \mathcal{E}}_{\pi}    \right],   \\
M_{-+}=&\frac{4\pi e f_{\pi}}{N_c Q}  \frac{\sqrt{|t^{\prime}|}}{2m} \xi\widetilde{ \mathcal{E}}_{\pi}.
\end{align}
\label{amp-gk}
\end{subequations}
These TFFs $\widetilde{\mathcal{H}}_{\pi}$ and $\widetilde{\mathcal{E}}_{\pi}$ (collectively $\widetilde{\mathcal{F}}_{\pi}$) factorize into a convolution of a perturbative hard-scattering kernel with the twist-2 pion DA $\phi_{\pi}(u)$ and the axial-vector GPDs $\widetilde{H}(x, \xi, t)$ or $\widetilde{E}(x, \xi, t)$. This factorized structure for DV$\pi$P is shown schematically in Fig.~\ref{fig:mp-diagram}.

Complementary to the longitudinal cross section in \eqref{cro-gk}, 
the transverse single-spin asymmetry (TSSA) is also a leading-twist observable relevant to the transversely polarized proton targets, 
which originates from the interference of the helicity amplitudes in \eqref{amp-gk}~\cite{Goloskokov:2009ia}. 
At leading twist, this observable is defined as~\cite{Frankfurt:1999fp,Frankfurt:1999xe,Diehl:2007hd}
\begin{align}
A_{\mathrm{UT}} &= \frac{\xi \sqrt{1-\xi^2} \texttt{Im}(\widetilde{\mathcal{H}}_{\pi} \widetilde{\mathcal{E}}_{\pi}^*)}{(1-\xi^2) \left| \widetilde{\mathcal{H}}_{\pi} \right|^{2} - \frac{t}{4m^2} \xi^2 \left| \widetilde{\mathcal{E}}_{\pi} \right|^{2} - 2 \xi^2 \texttt{Re}(\widetilde{\mathcal{H}}_{\pi} \widetilde{\mathcal{E}}_{\pi}^*)} \notag \\
&\quad \times \left( -\frac{\sqrt{|t'|}}{m} \right).
\label{eq:tsa}
\end{align}
The TSSA provides direct access to the relative phase between the TFFs $\widetilde{\mathcal{H}}_{\pi}$ and $\widetilde{\mathcal{E}}_{\pi}$,
offering unique sensitivity to the poorly constrained GPD $\widetilde{E}(x, \xi, t)$.

\vspace{0.2cm}

\noindent{\color{blue}\it Factorization formula for $\pi^+$ TFF.}
We start with the $\pi^+$ production.  In the collinear factorization regime, the TFF $\widetilde{ \mathcal{F}}_{\pi^+} $
admits the following factorized form at leading twist:
\beq
\label{ampl}
\widetilde{ \mathcal{F}}_{\pi^+}=\int^{1}_{-1} \mathrm{d}x \int^1_0 \mathrm{d}u\; \phi_{\pi}(u)\,T_{\pi^+}(u,x,\xi)\,\widetilde{F}^3(x,\xi,t),
\eeq
where  $\widetilde{F}^3(x,\xi,t)=\widetilde{F}^u(x,\xi,t)-\widetilde{F}^d(x,\xi,t)$ (with $\widetilde{F} 
\in \{\widetilde{H}, \widetilde{E}  \}$) signifies the the isovector combination of the quark GPDs of nucleon~\cite{Mankiewicz:1997uy}.
For brevity, the renormalization scale $\mu_R$ and the factorization scales $\mu_F$ 
in GPD and DA have been suppressed in \eqref{ampl} and thereafter.
It is worth remarking that the factorization theorem \eqref{ampl} has been proven to hold for all order of $\alpha_s$~\cite{Collins:1996fb}.

\begin{figure}
\centering
\includegraphics[width=0.5\textwidth]{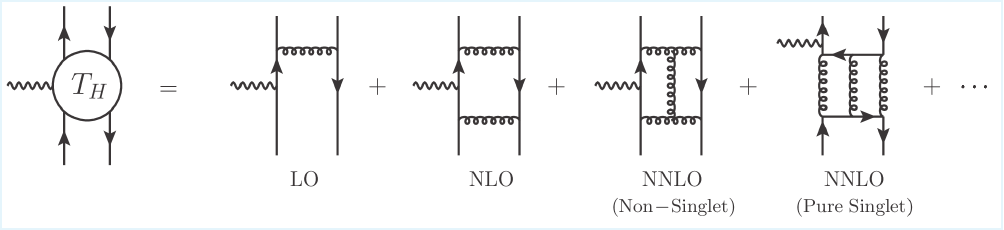}
\caption{Some representative parton-level diagrams for the deeply virtual $\pi$ production through two-loop order. 
Note that the rightmost two-loop diagram signifies the pure singlet contribution for $\pi^0$ leptoproduction.}
\label{Representative:Feynman:Diagrams}
\end{figure}

The hard-scattering kernel can be expressed in terms of 
the basic building block:
\beq
\label{amp2}
T_{\pi^+}(u,x,\xi)=  e_u T(\bar{u},-x,\xi)-e_d T(u,x,\xi),
\eeq
with $\bar{u}\equiv 1-u$.
Owing to asymptotic freedom of QCD, the coefficient function $ T(u,x,\xi)$ can be computed in perturbation theory order by order 
in $\alpha_s$,
\begin{align}
T(u, x,\xi)=C_F \alpha_s & \Big[  T^{(0)}(u, x,\xi)+\frac{\alpha_s}{\pi}T^{(1)}(u, x,\xi)
\nn\\
& +\left(\frac{\alpha_s}{\pi}\right)^2T^{(2)}(u, x,\xi)+\cdots \Big].
\label{eq:hadk}
\end{align}

The hard-scattering kernel can be determined through the perturbative matching procedure, {\it e.g.}, 
by computing the non-singlet (NS) partonic reaction $\gamma^{\ast} + u((x+\xi)\bar{p}) \bar{d}((\xi-x)\bar{p}) \to u(uk) \bar{d}(\bar{u}k)$
with $\bar{p}=(p+p')/2$ order by order in $\alpha_s$. 
Some representative Feynman diagrams through two-loop order are shown in Fig.~\ref{Representative:Feynman:Diagrams}.

Fortunately, it turns out that we do not need conduct the two-loop calculation 
from the scratch, instead we can directly infer the desired coefficient function
from the counterpart in the $\pi^+$ electromagnetic form factor (EMFF)
by making the following replacement:
\beq
T(u,x,\xi)=\frac{1}{2\xi}T_{\mathrm{EMFF}}\left(u,v\right)\bigg \vert_{v\to {x+(\xi-i\varepsilon)\over 2(\xi-i\varepsilon)}}.
\label{Replacement:formula}
\eeq
This shortcut works because in the Efremov-Radyushkin-Brodsky-Lepage (ERBL) region ($-\xi<x<\xi$), 
both the NS quark amplitudes and the evolution kernels~\cite{Belitsky:1999hf,Sarmadi:1982yg} in two processes are identical at any prescribed perturbative order,
so are the coefficient functions by iteratively solving the matching equations. 
The coefficient function for DV$\pi^+$P can then be analytically continued into the 
Dokshitzer-Gribov-Lipatov-Altarelli-Parisi (DGLAP) regime ($x>\xi$ or $x<-\xi$).
To circumvent the singularities located at $x=\pm \xi$, we utilize the causal $i\varepsilon$ prescription 
by making the substitution $\xi\xrightarrow{}\xi-i\varepsilon$ \cite{Duplancic:2016bge,Diehl:2007hd,Muller:2013jur}.

The  LO coefficient function $T^{(0)}(u,x,\xi)$ in $d=4-2\epsilon$ spacetime dimension reads~\cite{Vanderhaeghen:1999xj,Mankiewicz:1997uy,Frankfurt:1999xe,Frankfurt:1999fp,Mankiewicz:1998kg}
\beq
\label{amp4}
T^{(0)}(u, x,\xi)=\frac{1}{u(x+\xi-i \varepsilon)} (1-\epsilon).  
\eeq
The NLO coefficient function $T^{(1)}(u,x,\xi)$ has been deduced from the counterpart in the
NLO correction to $\pi^+$ EMFF long ago~\cite{Belitsky:2001nq,Diehl:2007hd}.
Consequently, following the recipe specified in \eqref{Replacement:formula}, we obtain 
the intended NNLO coefficient function, $T^{(2)}(u, x,\xi)$, 
from the recently available NNLO perturbative correction to $\pi^+$ EMFF~\cite{Chen:2023byr,Ji:2024iak}.

\vspace{0.2cm}

\noindent{\color{blue}\it Factorization formula for $\pi^0$ TFF.}
In the generalized Bjorken limit,  the TFF $\widetilde{\mathcal{F}}_{\pi^0}$ 
for DV$\pi^0$P can be expressed in terms of the sum of the non-singlet and pure singlet (PS) 
contributions~\footnote{In addition to the quark GPDs considered in this work, 
it has recently been suggested that the quark generalized transverse momentum-dependent distributions can also 
be probed through the azimuthal angular correlation in the same exclusive $\pi^0$ leptoproduction channel by including higher-twist contributions~\cite{Bhattacharya:2023hbq}.}:
\begin{align}
\label{pi0DVMP}
    \widetilde{\mathcal{F}}_{\pi^0}=&\int^{1}_{-1} \mathrm{d}x \int^1_0 \mathrm{d}u\; 
    \phi_{\pi}(u)\left[T_{\pi^0}^{\mathrm{NS}}(u,x,\xi)\widetilde{F}_{\pi^0}(x,\xi,t)\right.\notag\\+&\left.
    T_{\pi^0}^{\mathrm{PS}}(u,x,\xi)\widetilde{F}^{\mathrm{S}}(x,\xi,t)\right].
\end{align}
The NS channel contribution is intimately connected with DV$\pi^+$P:
\begin{subequations}
\begin{align}
T_{\pi^0}^{\mathrm{NS}}(u,x,\xi) = & T(\bar{u},-x,\xi)- T(u,x,\xi),   
\label{Coef:Function:DVPi0P}\\
\widetilde{F}_{\pi^0}(x,\xi,t)= & \tfrac{1}{\sqrt{2}} \left [  e_u \widetilde{F}^{u(-)}(x,\xi,t)- e_d\widetilde{F}^{d(-)}(x,\xi,t)\right],
\end{align}
\end{subequations}
with
\beq
\label{codd-gpd}
\widetilde{F}^{q(-)}(x,\xi,t)\equiv \widetilde{F}^{q}(x,\xi,t)-\widetilde{F}^{q}(-x,\xi,t),
\eeq
which specifically probes the C-odd combinations of nucleon GPDs~\cite{Muller:2013jur}.
Since the sea quark distributions cancel within the $C$-odd axial-vector GPDs, consequently we only need 
consider the valence quark contributions in the NS channel. 

The coefficient function $T$ in \eqref{Coef:Function:DVPi0P} is identical to that introduced in 
\eqref{eq:hadk} for DV$\pi^+$P, which has already been deduced from the higher order corrections to 
$\pi^+$ EMFF by employing \eqref{Replacement:formula}.

An interesting new type of contribution for DV$\pi^0$P stems from the singlet quark channel.
The corresponding singlet GPD in \eqref{pi0DVMP} is defined by 
\beq
\widetilde{F}^{\mathrm{S}}(x,\xi,t)=  
\widetilde{F}^{u(-)}(x,\xi,t)+\widetilde{F}^{d(-)}(x,\xi,t).
\label{Isospin:Singlet:GPD}
\eeq

Due to color and $C$-parity conservation, there receive no tree-level and one-loop contributions to the hard-scattering kernel in the PS channel. As represented in the rightmost diagram in Fig.~\ref{Representative:Feynman:Diagrams}, there arises a nonvanishing contribution starting at two-loop order. 
One of the major technical challenge of this work is to compute $T_{\pi^0}^{\mathrm{PS}}$ analytically.

\vspace{0.2cm}

\noindent{\color{blue}\it Description of the calculation.}
As aforementioned, the main challenge is to ascertain the two-loop PS hard-scattering kernel. 
We use the packages \texttt{QGRAF}~\cite{Nogueira:1991ex} and \texttt{HepLib}~\cite{Feng:2021kha} to generate Feynman diagrams and the corresponding amplitudes for the quark process $\gamma^*(q)+u(vl)\bar{u}(\bar{v}l) \to d(uk)\bar{d}(\bar{u}k)$.
There are in total 430 two-loop diagrams, one of which is depicted in the rightmost Feynman diagram of 
Fig.~\ref{Representative:Feynman:Diagrams}.
After employing \texttt{Apart}~\cite{Feng:2012iq} and \texttt{FIRE}~\cite{Smirnov:2019qkx} for partial fraction and 
integration-by-part reduction, we end up with 271 master integrals (MIs). 
They are calculated by using the method of differential equations~\cite{Kotikov:1990kg,Remiddi:1997ny,Gehrmann:1999as}, with the boundary conditions fixed by asymptotically expanding the MIs around $x=u=0$ with the aid of the method of region~\cite{Beneke:1997zp,Pak:2010pt,Jantzen:2011nz}. 
The boundary integrals are calculated by using the methods developed in Refs.~\cite{Chen:2019mqc,Chen:2023hmk}. 
The analytic expressions of all MIs are obtained by using the automated package \texttt{AmpRed}~\cite{Chen:2024xwt,Chen:2025paq}, which have also been numerically verified by \texttt{AMFlow}~\cite{Liu:2022chg}. 

It is worth stressing that, due to the presence of $\gamma_5$ in each quark line, 
the naive use of covariant projector approach~\cite{Chernyak:1983ej} in dimensional regularization would lead to 
incorrect result. 
We therefore give up taking the Dirac trace and keep all the spinor indices of external quarks open. 
By enforcing Ward-Takahashi identity to hold, we are able to obtain the UV and IR-finite result for $T_{\pi^0}^{\mathrm{PS}}$.
We devote Appendix~\ref{appendix:A} for a comprehensive description of our strategy to deduce $T_{\pi^0}^{\mathrm{PS}}$.

\vspace{0.2cm}

\noindent{\color{blue}\it Numerical Studies.}
The leading-twist pion DA is conveniently expanded in the Gegenbauer polynomial basis:
\begin{align}
\label{DAGegen}
\phi_\pi(u,\mu_F)=6u\bar{u}\sum_{n=0} a_{2n}(\mu_F)C_{2n}^{3/2}(2u-1),
\end{align}
where nonperturbative dynamics is encoded in the Gegenbauer moments $a_{2n}(\mu_F)$.
The recent NNLO study of pion EMFF~\cite{Chen:2023byr} suggests that the lattice QCD prediction by 
{\tt RQCD} Collaboration, 
$a_2(2\;\mathrm{GeV})=0.116^{+0.019}_{-0.020}$~\cite{RQCD:2019osh}, gives a better account of the data. 
Therefore we will adopt this value in our phenomenological exploration.

With the aid of the package \texttt{PolyLogTools}~\cite{Duhr:2019tlz}, the convolution between the hard-scattering kernel and pion DA over $u$
can be done analytically once the latter is expressed in the expanded form of \eqref{DAGegen}.
Thus the pion TFFs in \eqref{ampl} and \eqref{pi0DVMP} can be reduced into one-dimensional integrals,
which can be computed with high numerical precision by employing the numerical integrator \texttt{HCubature}~\cite{Cubature} with multiple-precision floating-point support. 

Although there have been impressive recent progress for the first-principle determination of nucleon GPDs from lattice QCD, 
so far the lattice predictions for nucleon axial-vector GPDs are available only for very limited values of $\xi$~\cite{Bhattacharya:2024wtg,Bhattacharya:2023jsc}. Therefore, we turn to two popular 
parameterizations of nucleon axial-vector GPDs for phenomenological purpose: Goloskokov-Kroll (GK) model~\cite{Goloskokov:2009ia,Goloskokov:2011rd,Kroll:2012sm,Goloskokov:2007nt} and 
the GPD through universal moment parametrization (GUMP) program~\cite{Guo:2025muf}. In the GK model, the isospin symmetry implies that the contributions from the 
$u$- and $d$-sea quarks cancel with each other inside $\widetilde{F}^3(x,\xi,t)$. 
Consequently, only the valence-quark GPDs contribute to DV$\pi^+$P~\cite{Belitsky:2001nq,Goloskokov:2011rd}. 
On the other hand, the {\tt GUMP 1.0} parametrization represents the first global extraction to consistently synthesize experimental data and lattice QCD constraints at NLO accuracy in the conformal moment expansion framework. 

In the numerical analysis, we fix $\mu_R=\mu_F=\mu$ and $n_L=3$. We use 
the package \texttt{FAPT}~\cite{Bakulev:2012sm} to evaluate the running QCD coupling constant to three-loop accuracy.
Three-loop evolution effect is incorporated in pion DA~\cite{Braun:2017cih,Strohmaier2018}.
We also use the package \texttt{tiktaalik}~\cite{Freese:2024ypk} to evolve the GPD from 2 GeV to any intended scale with 
two-loop accuracy.

\begin{figure}[htb]
    \centering
    \begin{subfigure}[h]{0.45\linewidth}
        \includegraphics[width=\linewidth]{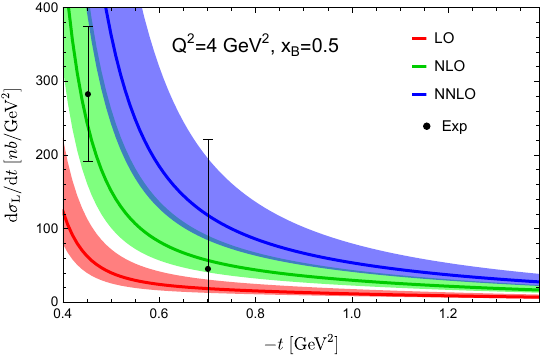}
    \caption{\texttt{JLab}, GK+{\tt RQCD}}   
    \end{subfigure}
    \begin{subfigure}[h]{0.45\linewidth}
        \includegraphics[width=\linewidth]{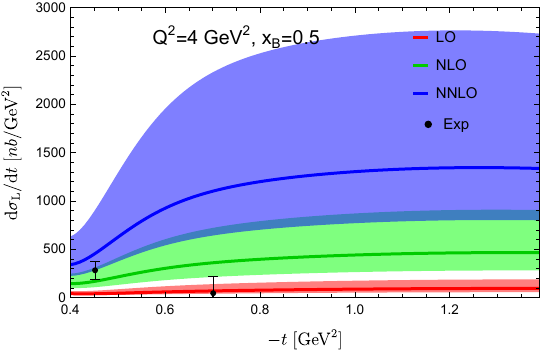}
    \caption{\texttt{JLab}, {\tt GUMP}+{\tt RQCD}}  
    \end{subfigure}
    \begin{subfigure}[h]{0.45\linewidth}
        \includegraphics[width=\linewidth]{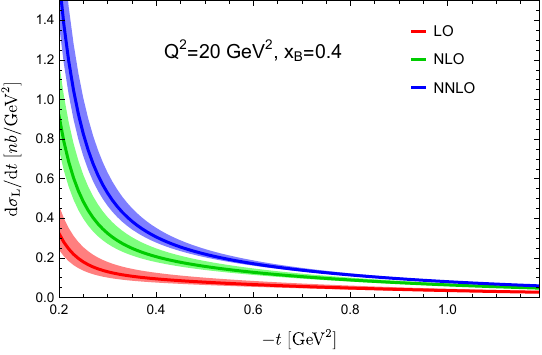}  
      \caption{\texttt{EicC}, GK+{\tt RQCD}} 
    \end{subfigure}
        \begin{subfigure}[h]{0.45\linewidth}
        \includegraphics[width=\linewidth]{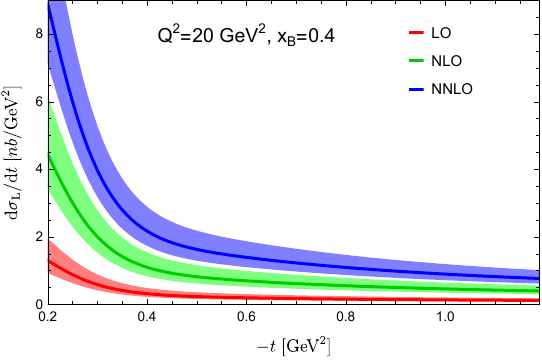}
       \caption{\texttt{EicC},  {\tt GUMP}+{\tt RQCD}}  
    \end{subfigure}
    \begin{subfigure}[h]{0.45\linewidth}
        \includegraphics[width=\linewidth]{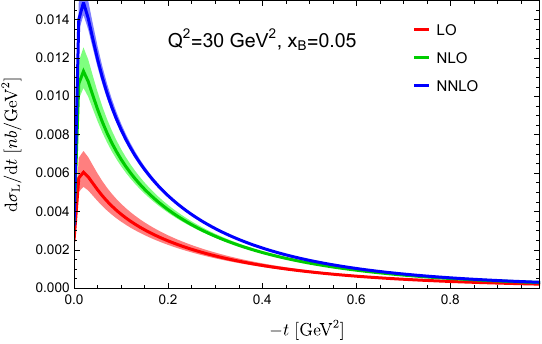}
        \caption{\texttt{EIC}, GK+{\tt RQCD}}
    \end{subfigure}
    \begin{subfigure}[h]{0.45\linewidth}
        \includegraphics[width=\linewidth]{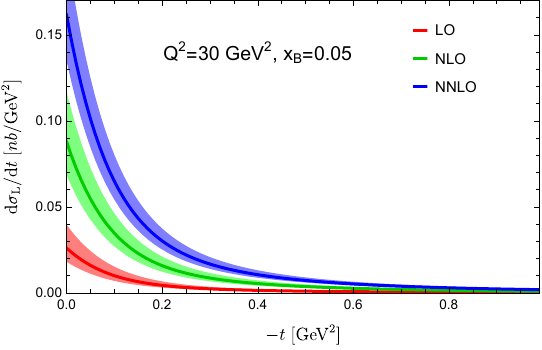}
        \caption{\texttt{EIC}, {\tt GUMP}+{\tt RQCD}}    
    \end{subfigure}
        \caption{The predicted $\mathrm{d}\sigma_{\mathrm{L}}/\mathrm{d}t$ for DV$\pi^+$P at various level of perturbative accuracy. The experimental data come from~\cite{Horn:2007ug}.}
    \label{fig:lc01r}
\end{figure}

We present the leading-twist predictions to DV$\pi^+$P at various perturbative order in Fig.~\ref{fig:lc01r}, 
choosing three benchmark kinematic points typical of the {\tt JLab}  ($Q^2=4$ GeV$^2$, $x_B=0.5$; upper panel), 
{\tt EicC}  ($Q^2=20$ GeV$^2$, $x_B=0.4$; middle panel), and {\tt EIC}  ($Q^2=30$ GeV$^2$, $x_B=0.05$; lower panel) facilities. 
The nonperturbative inputs are chosen as the GK+{\tt RQCD} (left column) and {\tt GUMP}+{\tt RQCD} (right column).
The uncertainties are estimated by varying the factorization and renormalization scales 
$\mu$ in the range $Q/\sqrt{2} \leq  \mu \leq \sqrt{2}Q$. 

As can be clearly visualized in Fig.~\ref{fig:lc01r},  
the NNLO perturbative correction to DV$\pi^+$P is positive and substantial, which is even 
as significant as the NLO correction at {\tt JLab} kinematic range~\cite{Horn:2007ug}.
Interestingly, the relative importance of the NNLO correction appears to decrease as $Q^2$ increases.
The perturbative convergence behavior, especially within the GK model, appears to be satisfactory at {\tt EIC} energy.

Within the GK model parametrization, the dominant contribution of $\pi^+$ leptoproduction 
stems from the pion pole term in the
$\widetilde{E}(x,\xi,t)$ term. As shown in the upper panel of Fig.~\ref{fig:lc01r},  
including the NNLO perturbative correction leads to an improved description of the measured longitudinal cross section~\cite{Horn:2007ug}, though the theoretical uncertainties remain sizable.
However, within the {\tt GUMP} parametrization, the NNLO prediction seems to fail to capture the measured $t$-dependence of the $\sigma_\mathrm{L}$. This may be partly attributed to the fact that 
the truncation procedure of $\xi$ implemented by {\tt GUMP} may not work well for $\xi=0.33$~\cite{Horn:2007ug}.

\begin{figure}[htb]
    \centering
    \begin{subfigure}[h]{0.45\linewidth}
        \includegraphics[width=\linewidth]{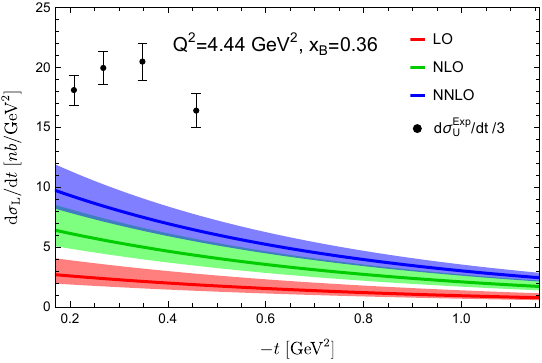}
        \caption{\texttt{JLab}, GK+{\tt RQCD}}
        \label{fig:pi0xbSr}
        
    \end{subfigure}
    \begin{subfigure}[h]{0.45\linewidth}
        \includegraphics[width=\linewidth]{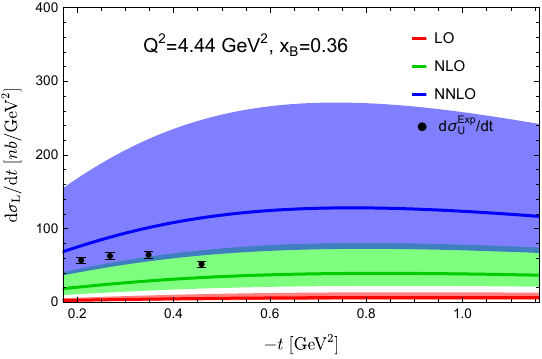}
        \caption{\texttt{JLab}, {\tt GUMP}+{\tt RQCD}}
        \label{fig:pi0xbSlpc}
        
    \end{subfigure}
     \begin{subfigure}[h]{0.45\linewidth}
        \includegraphics[width=\linewidth]{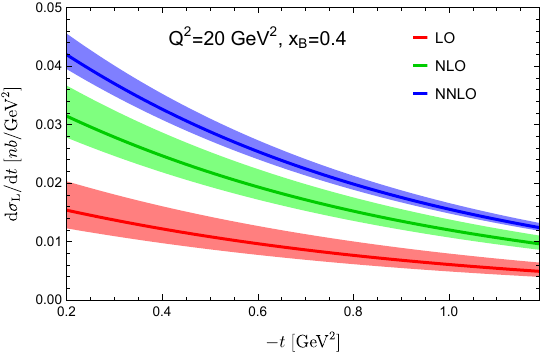}
        \caption{\texttt{EicC}, GK+{\tt RQCD}}
    \end{subfigure}
        \begin{subfigure}[h]{0.45\linewidth}
        \includegraphics[width=\linewidth]{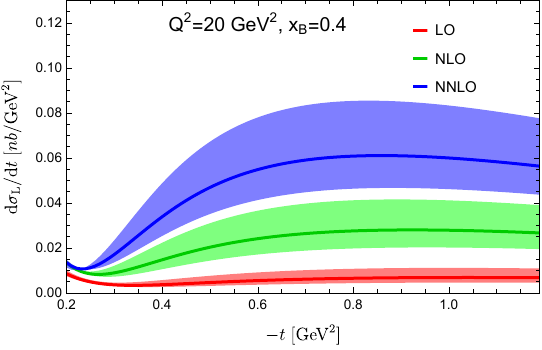}
        \caption{\texttt{EicC}, {\tt GUMP}+{\tt RQCD}}
    \end{subfigure}
     \begin{subfigure}[h]{0.45\linewidth}
        \includegraphics[width=\linewidth]{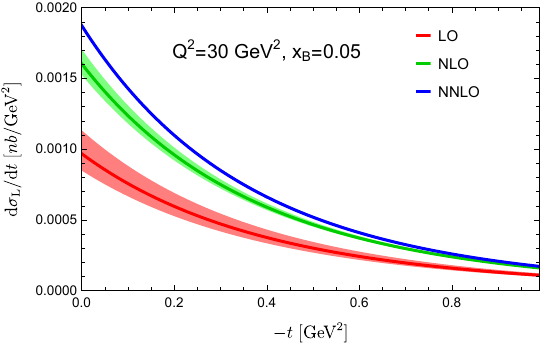}
        \caption{\texttt{EIC}, GK+{\tt RQCD}}
    \end{subfigure}
    \begin{subfigure}[h]{0.45\linewidth}
        \includegraphics[width=\linewidth]{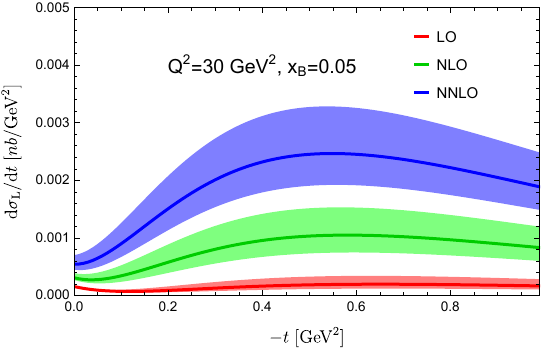}
        \caption{\texttt{EIC}, {\tt GUMP}+{\tt RQCD}}   
    \end{subfigure}
   \caption{The predicted $\mathrm{d}\sigma_{\mathrm{L}}/\mathrm{d}t$ for DV$\pi^0$P at various level of perturbative accuracy. The experimental data come from~\cite{JeffersonLabHallA:2020dhq}.   }
    \label{fig:p0a}
\end{figure}

Fig.~\ref{fig:p0a} displays the predicted $\mathrm{d}\sigma_{\mathrm{L}}/\mathrm{d}t$ for DV$\pi^0$P at various levels of 
perturbative accuracy, again taking the sample kinematics at {\tt JLab}, {\tt EicC} and {\tt EIC}.
Similar to $\pi^+$ leptoproduction, the NNLO correction is also positive and significant for DV$\pi^0$P, whose relative importance
with respect to the NLO correction appears to decrease with the increasing $Q^2$.

Due to the experimental incapability of separating the longitudinal DV$\pi^0$P cross section, 
we compare our predictions with the measured unpolarized cross section $\mathrm{d}\sigma_{\text{U}}/\mathrm{d}t = \mathrm{d}\sigma_{\text{T}}/\mathrm{d}t + \epsilon \mathrm{d}\sigma_{\text{L}}/\mathrm{d}t$~\cite{JeffersonLabHallA:2020dhq}, which is juxtaposed in the upper panel of Fig.~\ref{fig:p0a}. Where $\epsilon$ is the ratio of longitudinal and transverse photon flux.

However, even after including the NNLO corrections, the GK model predictions are still 
significantly lower than the measured $\sigma_\mathrm{U}$ (note the data points have been shrunk by a factor of 3 
for the sake of visibility). 
In contrast, the NLO and NNLO predictions based on the {\tt GUMP} parametrization appear 
to be compatible with the data. 

\begin{figure}[h]
    \centering
    \begin{subfigure}[t]{0.45\linewidth}
        \includegraphics[width=\linewidth]{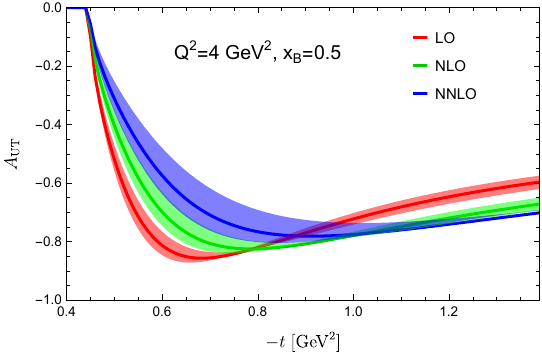}
        \caption{\texttt{JLab}, GK+{\tt RQCD}}
        \label{autjlab4rqcd}
    \end{subfigure}
    \begin{subfigure}[t]{0.45\linewidth}
        \includegraphics[width=\linewidth]{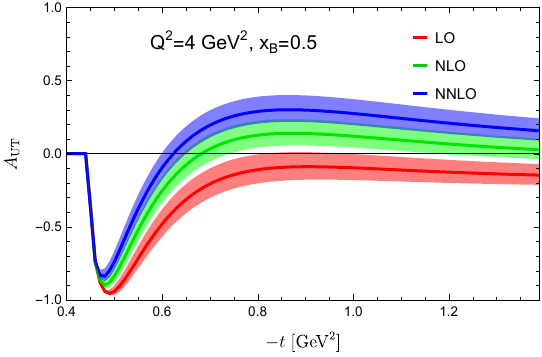}
        \caption{\texttt{JLab}, {\tt GUMP}+{\tt RQCD}}
        \label{autjlab4lpc}
    \end{subfigure}
   \begin{subfigure}[t]{0.45\linewidth}
        \includegraphics[width=\linewidth]{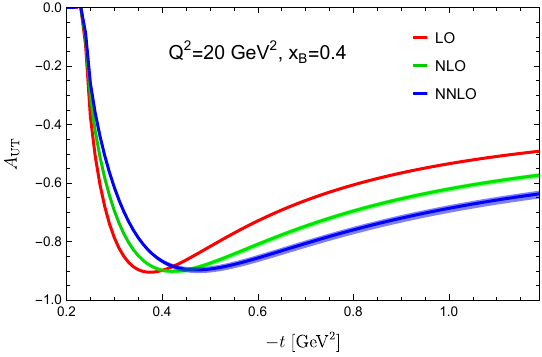}
        \caption{\texttt{EicC}, GK+{\tt RQCD}}
        \label{auteicc20ud}
    \end{subfigure}
   \begin{subfigure}[t]{0.45\linewidth}
        \includegraphics[width=\linewidth]{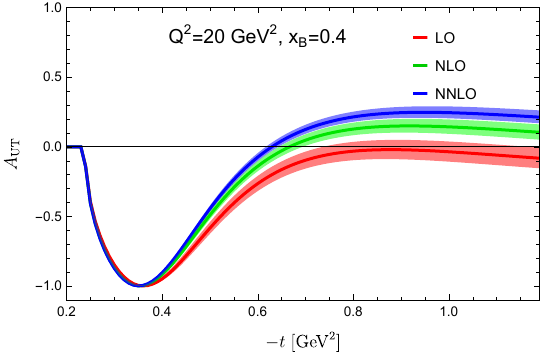}
        \caption{\texttt{EicC}, {\tt GUMP}+{\tt RQCD}}
        \label{auteicc20ud}
    \end{subfigure}
        \begin{subfigure}[t]{0.45\linewidth}
        \includegraphics[width=\linewidth]{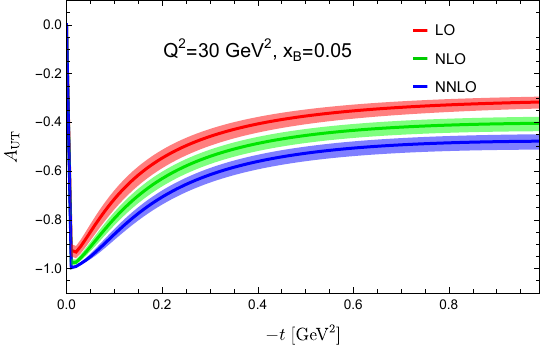}
        \caption{\texttt{EIC}, GK+{\tt RQCD}}
        \label{auteicrqcd}
    \end{subfigure}
     \begin{subfigure}[t]{0.45\linewidth}
        \includegraphics[width=\linewidth]{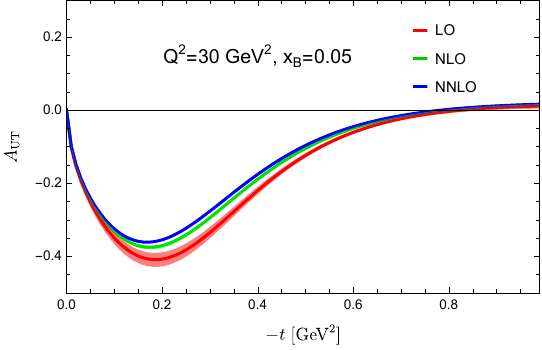}
        \caption{\texttt{EIC}, {\tt GUMP}+{\tt RQCD}}
        \label{auteiclpc}
    \end{subfigure}
    \caption{The predicted TSSA for DV$\pi^+$P at various level of perturbative accuracy. }
    \label{fig:tsa}
\end{figure}

In Fig.~\ref{fig:tsa} we also predict the $t$-dependence of the TSSA for DV$\pi^+$P at various level of
perturbative order.
Here we select representative $Q^2$ and $x_B$ values that span the kinematic reach of \texttt{JLab}, \texttt{EicC}, and \texttt{EIC}, 
mirroring the preceding longitudinal cross section analysis. We find that including 
NNLO corrections does not significantly alter the TSSA, echoing the conclusions drawn for NLO corrections in Refs.~\cite{Belitsky:2001nq,Diehl:2007hd}. Nevertheless, 
our results demonstrate that the TSSA remains sizable across this broad $Q^2$ range, a feature that is robust across different GPD parametrizations. 

Interestingly, the predicted TSSA at low $t$ and small $x_B$ diverges significantly between the GK and {\tt GUMP} models, 
as shown in the lower panel of Fig.~\ref{fig:tsa}. This discrepancy might originate 
from the dominant pion pole contribution from the $\widetilde{E}(x, \xi, t)$ term in the GK model, 
which is absent in the {\tt GUMP} parametrization. The pion pole dictates a $1/t$ enhancement at small $t$~\cite{Bechler:2009me}, 
generating a pronounced asymmetry. Consequently, future high-precision measurements of the TSSA at low $t$ will serve 
a crucial probe to test the picture of the pion-pole dominance.

\vspace{0.1 cm}
\begin{figure}[htb]
    \centering
    \begin{subfigure}[t]{0.45\linewidth}
        \includegraphics[width=\linewidth]{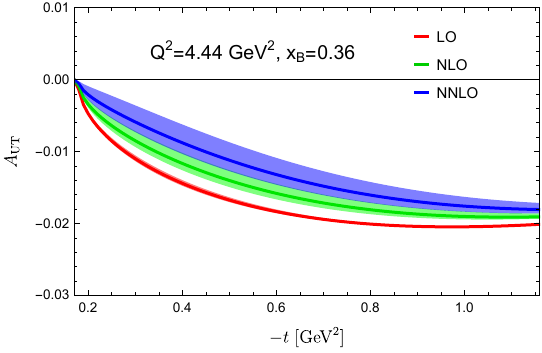}
        \caption{\texttt{JLab}, GK+{\tt RQCD}}
        \label{autjlab3rqcd}
    \end{subfigure}
    \begin{subfigure}[t]{0.45\linewidth}
        \includegraphics[width=\linewidth]{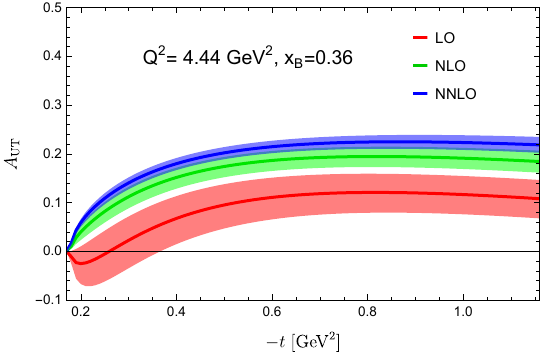}
        \caption{\texttt{JLab}, {\tt GUMP}+{\tt RQCD}}
        \label{autjlab3lpc}
    \end{subfigure}
   \begin{subfigure}[t]{0.45\linewidth}
        \includegraphics[width=\linewidth]{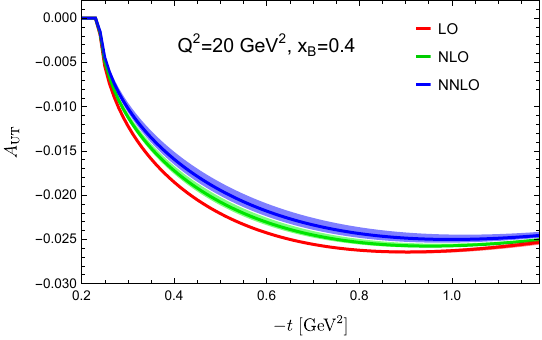}
        \caption{\texttt{EicC}, GK+{\tt RQCD}}
        \label{auteicc20ud}
    \end{subfigure}
    \begin{subfigure}[t]{0.45\linewidth}
        \includegraphics[width=\linewidth]{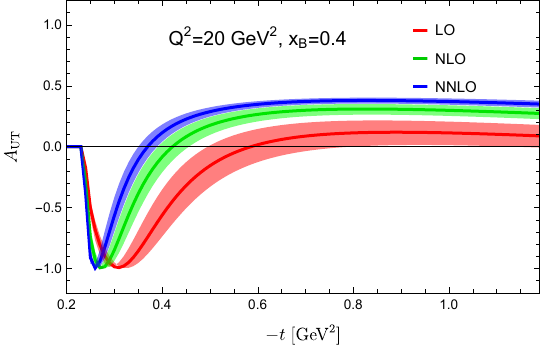}
        \caption{\texttt{EicC}, {\tt GUMP}+{\tt RQCD}}
        \label{auteicc20ud}
    \end{subfigure}
   \begin{subfigure}[t]{0.45\linewidth}
        \includegraphics[width=\linewidth]{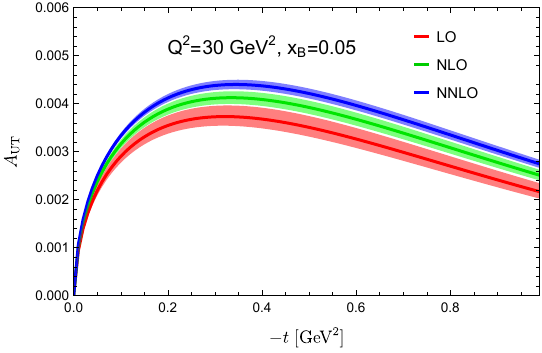}
        \caption{\texttt{EIC}, GK+{\tt RQCD}}
        \label{autp0eicrqcd}
    \end{subfigure}
    \begin{subfigure}[t]{0.45\linewidth}
        \includegraphics[width=\linewidth]{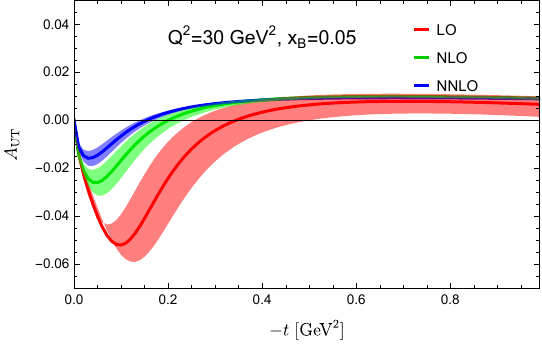}
        \caption{\texttt{EIC}, {\tt GUMP}+{\tt RQCD}}
        \label{autp0eiclpc}
    \end{subfigure}
    \caption{The predicted TSSA for DV$\pi^0$P at various level of perturbative accuracy.}
    \label{fig:tsa0}
\end{figure}

In Fig.~\ref{fig:tsa0}, we predict the TSSA for DV$\pi^0$P at various perturbative order, 
adopting the same kinematic setup as in Fig.~\ref{fig:tsa}. 
To align with the $\pi^+$ case, including higher order radiative corrections does not significantly alter the asymmetry. 
Due to the absence of the pion pole contribution, the $\pi^0$ TSSA predicted by the GK model (left panel) is notably small, posing a significant challenge for future experimental measurements. Conversely, the {\tt GUMP}
parametrization (right panel) yields a sizable asymmetry, but only at intermediate value of $x_B$.

\vspace{0.2cm}

\noindent{\color{blue}\it Summary.}  
In this work, we have presented the first calculation of NNLO QCD corrections to the
deeply virtual pion production ($\gamma^*_L N \to \pi N$) at leading twist in the collinear factorization approach. 
By relating the hard-scattering kernel in the non-singlet channel of the DV$\pi$P to that of the $\pi^+$ EMFF, 
and employing modern multi-loop computational techniques, we have derived the complete two-loop coefficient functions 
for both charged and neutral pion production channels. For the DV$\pi^0$P process, we have also 
accomplished the first calculation of the pure singlet contribution, whose phenomenological impact appears to be modest.

Employing two influential GPD parametrization models, we find that the NNLO perturbative corrections are positive and 
substantial. For typical kinematics setup at \texttt{JLab}, \texttt{EicC}, and the \texttt{EIC}, 
the inclusion of NNLO effects significantly enhances the longitudinal cross section compared to NLO predictions, 
even reaching more than a $100\%$ correction in the intermediate $Q^2$ region. 
These higher-order contributions considerably improve the agreement between 
perturbative QCD predictions and the available \texttt{JLab} data for DV$\pi^+$P. Furthermore, 
we have also investigated the transverse single-spin asymmetry $A_{\mathrm{UT}}$ for both $\pi^+$ and $\pi^0$ 
leptoproduction channels. While the magnitude of TSSA appears stable against the inclusion of the NNLO QCD correction, 
the $t$-dependence of TSSA exhibits a strong sensitivity to the underlying 
GPD parametrizations (notably the pion-pole contribution in the GK model). 

Our findings underscore the critical necessity of NNLO accuracy for a reliable extraction of nucleon GPDs from the vast amount of DVMP data expected at current and upcoming facilities. Furthermore, our work inaugurates a new era for exploring the effects of NNLO QCD corrections across a broader class of DVMP processes, including neutrino-induced meson production ($W^{\pm}+N \to M+ N^{\prime}$) and vector meson leptoproduction ($\gamma^*+N \to V+N'$, where $V=\rho, \omega, \phi, J/\psi, \Upsilon, \dots$).

\vspace{0.2 cm}

\begin{acknowledgments}
We acknowledge Yuxun Guo, Siwei Hu, Rong Wang and Yaping Xie for helpful discussions.
We are also indebted to Bernard Pire and Jian Zhou for valuable comments on the manuscript.
The work of W.~C. is supported in part by the NNSFC under Grant No.~11975200.
The work of F.~F. is supported in part by the NNSFC under Grant No.~12275353.
The work of Y.~J., G.~T. and Z.~Y.~W. is supported in part by the NNSFC under Grant No.~12475090.
The work of Q.~T.~S. is supported in part by the NNSFC under Grant No. ~12005191 and by the Natural Science Foundation of Henan Province under Grant  No. ~252300423011.
\end{acknowledgments}

\appendix
\begin{widetext}
\section{Calculation of the pure singlet contribution for DV$\pi^0$P}
\label{appendix:A}

As aforementioned in the main text, due to the presence of $\gamma_5$ in the quark line, 
some technical nuisance is encountered in projecting out the pure singlet (PS) contribution to DV$\pi^0$P 
in dimensional regularization. 
The purpose of this Appendix is to elaborate on a consistent recipe to compute the two-loop PS contribution. 

Let us take the following partonic reaction as an explicit example:
\beq
\gamma^*(q)+u(vl)\bar{u}(\bar{v}l) \to d(uk) \bar{d}(\bar{u}k),
\label{gamma*:uubar:to:ddbar}
\eeq
where two distinct quark lines are connected via gluons, as represented by the rightmost sample diagram in Fig.~\ref{Representative:Feynman:Diagrams}.
Note that the one-gluon exchange (tree) diagram yields a vanishing contribution due to the color-singlet requirement.
 
Had we naively employed the covariant projection technique~\cite{Chernyak:1983ej}, we would end up 
with the two-loop PS amplitude which bears wrong Lorentz structure and also 
contains the unwanted divergences. 
Unsurprisingly, the origin of this symptom can be traced to the fact that 
taking a Dirac trace containing a single $\gamma^5$ is a rather subtle issue 
in dimensional regularization.

In passing we note that similar symptom has already been encountered in the previous 
calculation of NLO QCD correction to the DV$\eta $P process, 
where the ‘t Hooft-Veltman-Breitenlohner-Maison (HVBM) scheme was employed there~\cite{Duplancic:2016bge}.  
In this work, we adopt a simpler method to overcome the $\gamma_5$ problem. 
  
With that said, at the very beginning we refrain from taking the Dirac trace 
and retain all the spinor indices open, 
and expand the amplitude of \eqref{gamma*:uubar:to:ddbar} in terms of a set of 
linearly-independent Dirac matrix bases. 
The corresponding $\gamma^*+u \bar{u} \to d \bar{d}$ amplitude
can then be written as $\mathcal{A}=\varepsilon^\mu(q,\lambda)\,\mathcal{A}_\mu$ and
\beq
{\cal A}_\mu = \left[\sum_k {\cal C}_{k} \gamma^{[\mu_0\cdots\mu_k]}\otimes \gamma_{[\mu_0\cdots\mu_k]}\right]_\mu, 
\label{Definition:A_mu}
\eeq
where
\beq
\gamma^{[\mu_0\cdots\mu_m]} \otimes \gamma_{[\nu_0\cdots\nu_m]}  
    \equiv u_\alpha(vl) \bar{v}_\beta(\bar vl)  \bar{u}_\rho(uk)  {v}_\sigma(\bar uk) 
    \, (\gamma^{[\mu_0\cdots\mu_m]})_{\beta\alpha} \, (\gamma_{[\nu_0\cdots\nu_m]} )_{\rho\sigma},
\label{Def:Dirac:Gamma:basis}
\eeq
with $\alpha,\beta,\rho,\sigma$ signifying the Dirac spinor indices. 
Concretely speaking, $\gamma^{[\mu_0\cdots\mu_m]}$ in \eqref{Def:Dirac:Gamma:basis} signifies the shorthand for 
the totally anti-symmetrized product of $m+1$ Dirac $\gamma$-matrices: 
\begin{equation}
\gamma^{[\mu_0\cdots\mu_m]} \equiv \frac{1}{(m+1)!} \sum_\sigma (-1)^{{\rm Sgn}(\sigma)} \gamma^{\mu_{\sigma_0} } \cdots \gamma^{\mu_{\sigma_m} } \,.
\end{equation}
Note that the Lorentz index $\mu$ in \eqref{Definition:A_mu} may be affiliated with either the coefficient
${\cal C}_k$ or the anti-symmetrized product of Dirac $\gamma$-matrices.

It is natural to demand that the amplitude obeys the Ward-Takahashi identity (current conservation),
which holds at any prescribed order in $\alpha_s$. Expanding the amplitude by
${\cal A}_\mu = \frac{\alpha_s}{\pi} {\cal A}^{(1)}_\mu + \left(\frac{\alpha_s}{\pi}\right)^2 {\cal A}^{(2)}_\mu +\cdots$, we thus expect $q^\mu {\cal A}^{(1,2)}_\mu =0$.
 
Explicit manipulation over the one- and two-loop amplitudes yield
\begin{subequations}
\begin{align}
\label{eq:Amp1}
q^\mu{\cal A}_\mu^{(1)} &= {\cal C}^{(1)}_1\left\{ \gamma^{[k]} \otimes \gamma_{[l]} - l\cdot k\, \gamma^{[\mu]} \otimes \gamma_{[\mu]} \right\} 
\nn\\
&\quad + {\cal C}_2^{(1)}\left\{ 3 \gamma^{[k\mu_1\mu_2]} \otimes \gamma_{[l\mu_1\mu_2]} - l\cdot k\, \gamma^{[\mu_0\mu_1\mu_2]} \otimes \gamma_{[\mu_0\mu_1\mu_2]} \right\}, 
\\
q^\mu {\cal A}_\mu^{(2)} &= {\cal C}_1^{(2)}\left\{ \gamma^{[k]} \otimes \gamma_{[l]} -l\cdot k\, \gamma^{[\mu]} 
\otimes \gamma_{[\mu]} \right\} 
\nn\\
&\quad + {\cal C}_2^{(2)}\left\{ 3 \gamma^{[k\mu_1\mu_2]} \otimes \gamma_{[l\mu_1\mu_2]} - l\cdot k\, \gamma^{[\mu_0\mu_1\mu_2]} \otimes \gamma_{[\mu_0\mu_1\mu_2]} \right\} 
\nn\\
&\quad + {\cal C}_3^{(2)}\left\{ 5 \gamma^{[k\mu_1\mu_2\mu_3\mu_4]} \otimes \gamma_{[l\mu_1\mu_2\mu_3\mu_4]} - l\cdot k\, \gamma^{[\mu_0\mu_1\mu_2\mu_3\mu_4]} \otimes \gamma_{[\mu_0\mu_1\mu_2\mu_3\mu_4]} \right\} \, .
\label{eq:Amp2}
\end{align}
\label{qmu:Amu:one:and:two:loop}
\end{subequations}
To condense the notation, we often replace a Lorentz index $\mu_i$ of the 
totally anti-symmetrized product of Dirac matrices with the external momentum.
It should be understood as a contraction of the Lorentz index, say, 
$\gamma^{[p]}\equiv \cancel{p}$. 

Now let us state a key observation. As a sufficient condition for the right-hand sides of
\eqref{qmu:Amu:one:and:two:loop} to vanish, 
the following identity should hold true in $d$-dimensional spacetime:
\beq
\gamma^{[\mu_0\mu_1\cdots\mu_{2n}]} \otimes \gamma_{[\mu_0\mu_1\cdots\mu_{2n}]} = \frac{2n+1}{l\cdot k} \gamma^{[k\mu_1\cdots\mu_{2n}]} \otimes \gamma_{[l\mu_1\cdots\mu_{2n}]}.
\label{sufficient:condition}
\eeq
Here we attempt to present a proof to \eqref{sufficient:condition}. 
It is convenient to adopt the light-cone coordinate $a^\mu=(a^+,a^-,{\bf a}_\perp)$. 
Lorentz covariance requires that the Lorentz indices $\{\mu_i\}$ can enter $u_\alpha(vl) \bar{v}_\beta(\bar vl) \, (\gamma^{[\mu_0\cdots\mu_m]})_{\beta\alpha}$ only through the form $l^{\mu_i}$, 
$g^{\mu_i\mu_j}$ (which nevertheless 
is forbidden due to the anti-symmetric feature under $\mu_i\leftrightarrow\mu_j$ in $\gamma^{[\mu_0\cdots\mu_m]}$), $\varepsilon^{\mu_{i} \mu_{i_1} \mu_{i_2} \mu_{i_3}}$ or 
$\varepsilon^{\mu_{i}\mu_{i_1} \mu_{i_2} l}$ (Similar to 't Hooft-Veltman
prescription~\cite{tHooft:1972tcz}, here we also assume the totally-antisymmetric 
$\varepsilon$-tensor is a four-dimensional object). 
Therefore, one is forced to conclude that there should be at least one and only one $\mu_i$ index 
to align with the direction of $k$ (Recall that $k$ and $l$ are two conjugate light-like vectors and 
$k\cdot l= \, k^+\,l^- \neq 0$). 
Consequently, one can rewrite
\beq
u_\alpha(vl) \bar{v}_\beta(\bar vl) \, (\gamma^{[\mu_0\cdots\mu_m]})_{\beta\alpha} = 
 u_\alpha(vl) \bar{v}_\beta(\bar vl) \, (\gamma^{[\mu_0\cdots\mu_m]})_{\beta\alpha} \sum_{i=0}^m \delta^{\vert\vert}_{\mu_i k}.
\eeq
with $\delta^{\vert\vert}_{\mu_i k}=1$ for $\mu_i \vert\vert k$ and zero otherwise.

In a similar fashion, one can show that there should be at least one and only one $\nu_i$ index 
in $\bar{u}_\rho(uk)  {v}_\sigma(\bar uk) \, (\gamma_{[\nu_0\cdots\nu_m]} )_{\rho\sigma}$
to align with the direction of $l$.

We thus arrive at
\bqa
\gamma^{[\mu_0\mu_1\cdots\mu_{2n}]} \otimes \gamma_{[\mu_0\mu_1\cdots\mu_{2n}]} 
&=& 
\sum_{i=0}^{2n} \delta^{\vert\vert}_{\mu_i k}\delta^{\vert\vert}_{\nu_i l}\gamma^{[\mu_0\mu_1\cdots\mu_{2n}]} \otimes \gamma^{[\nu_0\nu_1\cdots\nu_{2n}]}g_{\mu_0\nu_0}\cdots g_{\mu_{2n}\nu_{2n}}
\nn\\
&=&
(2n+1)\delta^{\vert\vert}_{\mu_0 k}\delta^{\vert\vert}_{\nu_0 l} \gamma^{[\mu_0\mu_1\cdots\mu_{2n}]} \otimes \gamma^{[\nu_0\nu_1\cdots\nu_{2n}]}g_{\mu_0\nu_0}\cdots g_{\mu_{2n}\nu_{2n}}
\nn\\ 
&=& 
\frac{2n+1}{l\cdot k} \gamma^{[k\mu_1\cdots\mu_{2n}]} \otimes \gamma_{[l\mu_1\cdots\mu_{2n}]}.
\eqa

The current conservation indicates that ${\cal A}^\mu \propto (k+l)^\mu$. 
To facilitate the extraction of the intended PS hard kernel, it  
is convenient to further contract the $\mathcal{A}_\mu$ with $(k+l)^\mu$:
\begin{subequations}
\bqa
(k+l)^\mu{\cal A}^{(1)}_\mu &=& {\cal G}_1^{(1)} \gamma^{[k]} \otimes \gamma_{[l]}
+ {\cal G}_2^{(1)}\gamma^{[k\mu_1\mu_2]} \otimes \gamma_{[l\mu_1\mu_2]},
\\
(k+l)^\mu{\cal A}^{(2)}_\mu &=& {\cal G}_1^{(2)} \gamma^{[k]} \otimes \gamma_{[l]}
+ {\cal G}_2^{(2)}\gamma^{[k\mu_1\mu_2]} \otimes \gamma_{[l\mu_1\mu_2]}
+ {\cal G}_3^{(2)} \gamma^{[k\mu_1\mu_2\mu_3\mu_4]} \otimes \gamma_{[l\mu_1\mu_2\mu_3\mu_4]} \, .
\label{two:loop:contraction}
\eqa
\end{subequations}

After substituting the expressions of all the MIs, 
we find that all the divergences are contained in $\mathcal{G}^{(1,2)}_1$, at both one-loop and two-loop order. 
Fortunately, the accompanying spinor cluster $\gamma^{[k]} \otimes \gamma_{[l]}$ 
generates a vanishing result after projecting the final state quark-antiquark pair into 
the $\pi^0$ quantum number.

Since ${\cal G}^{(1,2)}_2$ and ${\cal G}^{(2)}_3$ are both UV and IR finite, 
from now on we can safely go back to $d=4$ dimension. 
For the $\gamma^*+u \bar{u} \to d \bar{d}$ process,
the virtual photon can be attached either to the initial state
quark line or final state quark line. 
The coefficient $\mathcal{G}$ can be decomposed as $\mathcal{G}=e_u \mathcal{G}^{\rm in}+e_d \mathcal{G}^{\rm out}$. 
However, in order to infer the complete PS hard kernel, we must also take into account 
the PS amplitude for the $\gamma^*+u \bar{u} \to u \bar{u}$ process, 
in line with the flavor superposition of neutral pion $\pi^0 =
{1\over \sqrt{2}}(u\bar{u}-d\bar{d})$. Therefore, the short-distance coefficient reads
${\mathcal G}^{\pi^0}= e_u \frac{1}{\sqrt{2}}({\mathcal G}^{\rm in}-{\mathcal G}^{\rm in})+
\frac{1}{\sqrt{2}}(e_u -e_d ){\mathcal G}^{\rm out}=\frac{1}{\sqrt{2}}{\mathcal G}^{\rm out}$.
We should also take into account the contributions of the partonic channels 
$\gamma^*+d \bar{d} \to d \bar{d}$ and $\gamma^*+d \bar{d} \to u \bar{u}$, whose effect is reflected
in the isospin-singlet combination of nucleon GPDs in \eqref{Isospin:Singlet:GPD}.
 
At one-loop level, it turns out that $\mathcal{G}_2^{{\rm out}(1)}=0$ 
due to the conservation of $C$ parity, 
since the $C$-even $\pi^0$ cannot couple to a virtual photon plus
two gluons. 

At two-loop order, the last term in \eqref{two:loop:contraction} contains
$\gamma^{[k\mu_1\mu_2\mu_3\mu_4]} \otimes \gamma_{[l\mu_1\mu_2\mu_3\mu_4]}$, 
which trivially vanishes in four spacetime dimension, thus can be dropped.
It turns out that only the ${\cal G}_2^{{\rm out}(2)}$ term survives.
Applying the covariant projection technique, 
we then obtain the desired two-loop PS hard-scattering kernel:
\beq
T^{\mathrm{PS}}_{\pi^0}=\left(\frac{\alpha_s}{\pi}\right)^2\frac{1}{8\xi Q^2} {\mathcal G}_2^{\pi^0(2)}\,\mathrm{Tr}[\cancel{l}\gamma^5\gamma^{[k\mu_1\mu_2]}] \,\mathrm{Tr}[\cancel{k}\gamma^5\gamma^{[l\mu_1\mu_2]}].
\eeq

\begin{figure}[t]
    \centering
    \begin{subfigure}[h]{0.45\linewidth}
        \includegraphics[width=\linewidth]{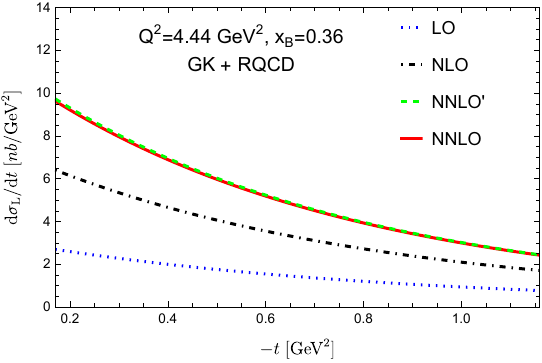}
    \end{subfigure}
    \begin{subfigure}[h]{0.45\linewidth}
        \includegraphics[width=\linewidth]{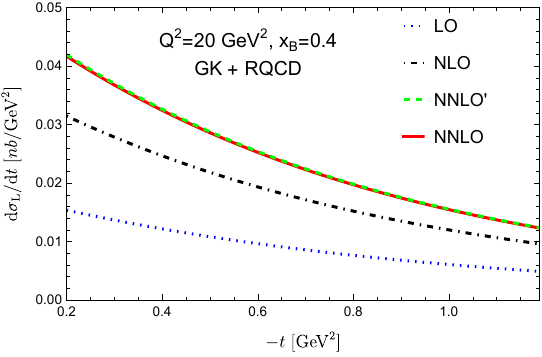}
    \end{subfigure}
    \caption{The predicted $\mathrm{d}\sigma_{\mathrm{L}}/\mathrm{d}t$ at various level of perturbative accuracy. 
   The curve labelled with NNLO' implies that only the two-loop NS contribution is considered, while the curve
   labelled with NNLO implies that both the NS and PS contributions are included. }
\label{fig:pScross}
\end{figure}

\begin{figure}[t]
    \centering
    \begin{subfigure}[h]{0.45\linewidth}
        \includegraphics[width=\linewidth]{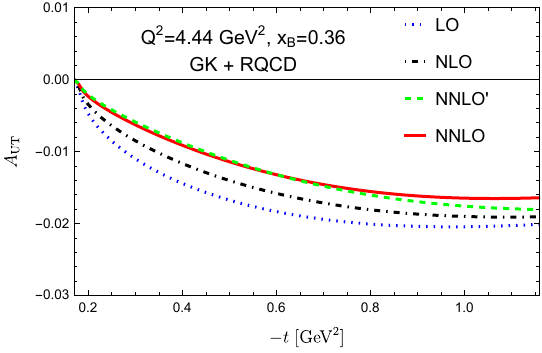}
    \end{subfigure}
    \begin{subfigure}[h]{0.45\linewidth}
        \includegraphics[width=\linewidth]{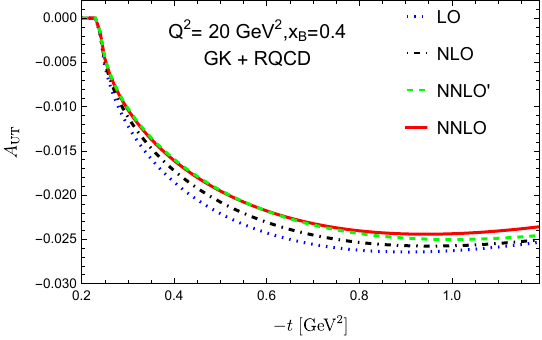}
    \end{subfigure}
    \caption{The predicted TSSA at various level of perturbative accuracy. The meaning of the labels NNLO' and NNLO
   is the same as in Fig~\ref{fig:pScross}.}
    \label{fig:pSAUT}
\end{figure}

It is worth stressing that,    
the phenomenological impact of the PS contribution for DV$\pi^0$P appears to be rather modest 
with respect to the two-loop NS contribution.   
As one readily observes from Fig.~\ref{fig:pScross}, including the PS contribution only decreases the NNLO prediction 
for ${\mathrm{d}}\sigma_L/{\mathrm{d}t}$ by about 1\%, in both {\tt JLab} and {\tt EicC} experiments. 
In comparison, the effect of the PS contribution for TSSA is somewhat more important. 
As shown in Fig.~\ref{fig:pSAUT}, one can see at large $|t|$ regime, including the PS contribution could change the NNLO predictions
for $A_{\rm UT}$ by about 10\%, in both {\tt JLab} and {\tt EicC} experiments. 

\end{widetext}


\end{document}